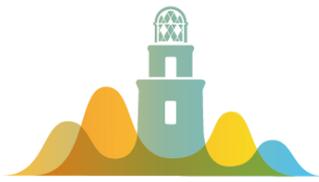

# TOWARDS PREDICTING BINAURAL AUDIO QUALITY IN LISTENERS WITH NORMAL AND IMPAIRED HEARING

**Thomas Biberger**[*]    **Stephan D. Ewert**

Medizinische Physik and Cluster of Excellence "Hearing4All", Universität Oldenburg, Germany

## ABSTRACT

Eurich et al. (2024) recently introduced the computationally efficient monaural and binaural audio quality model (eMoBi-Q). This model integrates both monaural and binaural auditory features and has been validated across six audio datasets encompassing quality ratings for music and speech, processed via algorithms commonly employed in modern hearing devices (e.g., acoustic transparency, feedback cancellation, and binaural beamforming) or presented via loudspeakers. In the current study, we expand eMoBi-Q to account for perceptual effects of sensorineural hearing loss (HL) on audio quality. For this, the model was extended by a nonlinear auditory filterbank. Given that altered loudness perception is a prevalent issue among listeners with hearing impairment, our goal is to incorporate loudness as a sub-dimension for predicting audio quality in both normal-hearing and hearing-impaired populations. While predicting loudness itself is important in the context of loudness-based hearing aid fitting, loudness as audio quality sub-measure may be helpful for the selection of reliable auditory features in hearing impaired listeners. The parameters of the filterbank and subsequent processing stages were informed by the physiologically-based (binaural) loudness model proposed by Pieper et al. (2018). This study presents and discusses the initial implementation of the extended binaural quality model.

_________________
[*]**Corresponding author**: thomas.biberger@uol.de.

## 1. INTRODUCTION

Audio or speech signals can be distorted by sound reproduction systems or hearing device algorithms until they reach the eardrum of the listeners. Therefore, for the perceptual evaluation of such distortions it is important to measure audio quality. Although measurements are still the "gold standard", they are typically time consuming, require lab facilities and in some cases expert listeners. Therefore, it is highly desirable to have reliable audio quality models as they offer fast predictions and are ready for use anywhere at any time. Given that devices and algorithms may introduce monaural and binaural distortions, audio quality measures have to consider both aspects. In this regard, Fleßner et al. [1] successfully combined the output of the monaural GPSM$^q$ [2] with the outputs of the binaural BAM-Q [3], resulting in the monaural and binaural model (MoBi-Q) to predict overall audio quality. Given, that MoBi-Q is based on the combination of two separate models, it is computationally not very efficient, and peripheral processing is calculated separately for monaural and binaural model stages. To overcome this, we suggested an efficient version of the MoBi-Q, termed eMoBi-Q [4], which uses a joint peripheral processing stage from which monaural and binaural features were derived and combined in the model back end for the prediction of overall audio quality. eMoBi-Q calculates differences in (monaural) spectral coloration, interaural level differences and the complex interaural correlation coefficient, between a reference and a test signal from which audio quality is predicted.

All previously mentioned models predict audio quality for normal-hearing (NH) listeners and do not capture consequences of sensorineural hearing on audio quality perception. The HASQI [5-6] and HAAQI [7] models can predict audio quality data from hearing-impaired (HI) listeners by using audiogram information to simulate consequences of peripheral hearing loss. However, both models only account for monaural audio quality and do not consider spatial distortions. Therefore, a binaural audio quality model for NH and HI listeners is highly desirable.

For both NH and HI listeners the loudness of a signal is an important perceptual attribute that is used in many different areas such as applied psychoacoustics or hearing aid fitting to improve speech understanding. Given that loudness perception alters with hearing loss, the integration of binaural loudness aspects seems a reasonable extension of the already existing features in eMoBi-Q. Recently, Pieper et al. [8-9] proposed a binaural loudness model applicable to NH listeners and for individualized loudness predictions of HI listeners. The peripheral processing in their model is simulated by a transmission-line model, which is computationally expensive, while their back end for binaural loudness calculation does not require much computational effort.

In this study, a first step towards a binaural audio quality model for NH and HI is made by replacing the linear peripheral processing stage in eMoBi-Q by the non-linear Gammawarp filterbank [10]. This filterbank allows to simulate consequences of peripheral hearing loss based on audiogram information. Here, the revised eMoBi-Q with Gammawarp filterbank is evaluated for four databases including monaural, binaural, and combined monaural and binaural distortions related to hearing devices and its prediction performance is compared to that of the original eMoBi-Q [4]. An additional research question was, whether the Gammwarp filterbank can be combined with the back end for binaural loudness calculation proposed by Pieper et al. [9], to provide reasonable loudness predictions within the framework of the suggested audio quality model. Three loudness experiments from literature considering different aspects were used for the evaluation.

## 2. AUDIO QUALITY MODEL

The block diagram of the proposed audio quality model is shown in Figure 1. Processing stages different to Eurich et al. [4] are highlighted in red. Audio quality predictions were based on monaural spectral coloration, complex interaural correlation coefficient γ, and interaural level cues. The binaural loudness features were separately evaluated with three loudness experiments and did not contribute to audio quality predictions reported in this study.

### 2.1 Peripheral processing

A middle-ear filtering was applied before signals are processed by the Gammawarp filterbank [10]. The outputs of the filterbank were 23 bandpass-filtered signals with center frequencies ranging from 80 Hz to 12500 Hz. For audio quality predictions, channels with center frequencies ranging from 315 Hz to 12500 Hz were used. In the next step hearing thresholds were simulated, so that signals with levels below a certain threshold level were not considered. Then, low-pass filtered envelope signals (Env) were calculated by applying a first-order low-pass filter with 150-Hz cutoff frequency to the Hilbert envelope. Env signals were used for the monaural spectral coloration feature, ILD feature, and γ feature for frequency bands above 1.3 kHz. Temporal fine structure (TFS) information was used for γ features below 1.3 kHz.

### 2.2 Feature extraction & combination for audio quality predictions

All features were calculated in consecutive time frames of 400 ms and combined identical to [4], which is briefly described in the following.

Monaural spectral coloration features were calculated based on power-based signal-to-noise ratios (SNRs) for the increment and decrement case, referring to the cases where the device under test either introduced or removed energy from the original signal. The increment and decrement SNRs, $SNR_{incr}(n,p)$ and $SNR_{decr}(n,p)$, were averaged across time frames $n$, resulting in $SNR_{incr}(p)$ and $SNR_{decr}(p)$ for each frequency channel $p$. Both increment and decrement SNRs were averaged for each frequency channel and the resulting mean values were optimally combined across frequency channels providing the single-valued $SNR_{mon}$. A logarithmic transformation with lower and upper bounds was applied to limit the range of the spectral coloration measure (see Eq. 10 in [4]) providing $d'_{mon}$.

For frequency channels below 1.3 kHz the complex correlation coefficient γ was calculated based on temporal fine structure, while above of 1.3 kHz it was calculated based on the Hilbert envelope:

$$\gamma(n, p) = \frac{\overline{l(n, p) \cdot r(n, p)}}{\sqrt{\overline{|l(n, p)|^2} \, \overline{|r(n, p)|^2}}} \qquad (1)$$

The overhead bars denote the mean over the duration of the 400-ms time frame. Interaural level differences were extracted as the logarithmic power ratio between the left and right signal.

The output of the binaural model path resulted from a weighted optimal combination of the sensitivity indices for γ- and ILD-features:

$$d'_{bin} = \sqrt{d'^2_\gamma + \alpha \cdot d'^2_{ILD}} \qquad (2)$$

Where $\alpha = 1/13$ refers to the weighting of ILD features as is was used in [4]. Then the perceptual range of $d'_{mon}$ and $d'_{bin}$ was normalized to a scale ranging from 0 to 1. Finally, overall quality was determined by the lower quality-component, either originating from the monaural or binaural part.

### 2.3 Binaural loudness predictions

The binaural loudness prediction is based on the outputs of the nonlinear Gammawarp filterbank for the left and right ear. Here frequency channels with center frequencies ranging from 80 Hz to 12500 Hz were used. The same calculation steps as shown in Fig. 1 of Pieper et al. [9] were applied in this study and are briefly described here: First, temporal integration was simulated by a first-order low-pass filter with an integration time of 25 ms, followed by a pre-attenuation, that mimics inner hair cell loss. Next, the absolute hearing threshold was simulated, followed by a post gain that represents a linear amplification applied to the signal part above the hearing threshold and is assumed to reflect effects of central gain in the auditory system. To be able to further adapt the model to individual loudness perception, a bandwidth-dependent gain was applied to the post gain outputs that weights the signals according to their bandwidth and accordingly effects spectral loudness summation for narrowband and broadband stimuli. Afterwards, the left and right ear signals were combined in the binaural summation stage that assumes binaural inhibition. The resulting model outputs, representing the specific loudness per frequency channel, were summed over frequency channels, resulting in a time series of binaural loudness values. The maximum value was taken to obtain the internal loudness. The internal loudness was then transformed to obtain the final binaural loudness value in sones. With exception to the post gains, the same parameters as suggested in [8-9] were used here.

### 3. EVALUATION

#### 3.1 Audio quality

The first database includes 60 diotic speech and music items processed by adaptive feedback cancelation algorithms and was taken from [11]. The second database includes 16 speech items processed with the speech-distortion-weighted multichannel filter and was taken from the evaluation in [3]. Such beamformer reduces the noise but also changes binaural cues. The third database [12] includes 32 speech signals processed by the binaural minimum-variance-distortionless-response (MVDR) algorithms to preserve the interaural coherence of diffuse noise fields. In the fourth database [13], 140 speech and music items were used for audio quality evaluation of a real-time hearing device prototype for acoustically transparent sound reproduction by applying feedback suppression and using a null-steering beamformer and individualized equalization of the sound pressure at the eardrum.

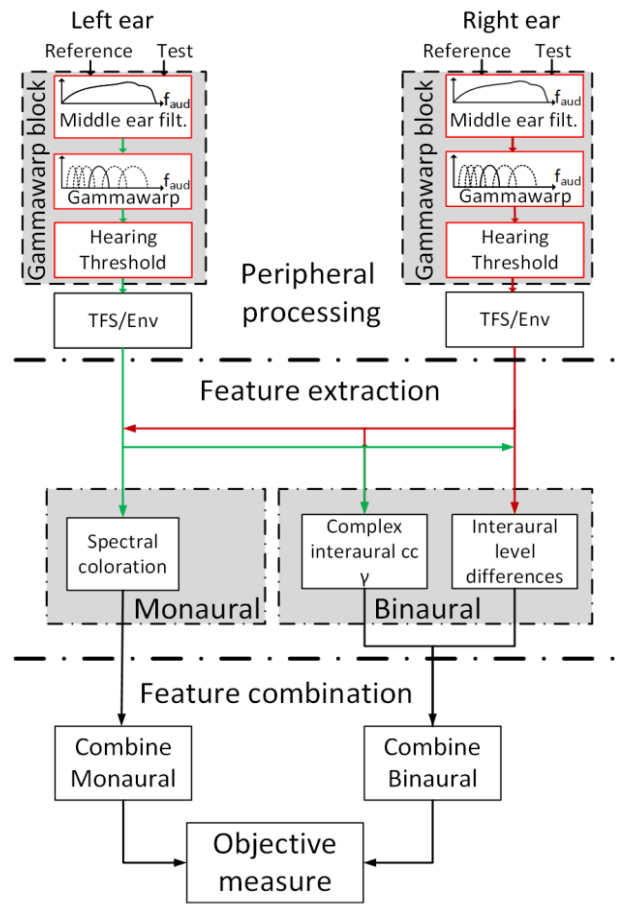

**Figure 1**. Block diagram of the proposed audio quality model with peripheral processing, feature extraction and feature combination. Processing stages different to [4] are highlighted in red. Binaural loudness features (not shown here) are based on the outputs of the Gammawarp filterbank and calculated according to [9].

Signals in the first database have monaural distortions, whereas signals in the second and third database show purely binaural distortions. The fourth database includes signals with monaural and binaural distortions.

### 3.2 Loudness

Three experiments considering different aspects of loudness perception were used in this study. Data for the first experiment on loudness as a function of level was taken from Table 1 in [14]. They used 500-ms, 1-kHz pure tones ranging from 0 to 100 dB sensation level, which were approximately equivalent to sound pressure levels (SPL) in a free sound field.

In the second experiment, data for equal loudness contours and absolute thresholds for 0 to 50 phones were taken from [15]. Signals were 400 ms in duration, and pure tones ranged from 100 Hz to 10 kHz.

In the third experiment, the model was tested for spectral loudness summation based on [16]. Signals were 1 s in duration, narrow-band Gaussian noise, and had bandwidths of 200 Hz, 400, Hz, 800 Hz, 1600 Hz, 3200 Hz, and 6400 Hz. The reference bandwidth was 3200 Hz presented at levels of 45, 55, and 65 dB SPL. All noise signals were geometrically centered at 2 kHz.

## 4. RESULTS AND DISCUSSION

The results of the audio quality evaluation provided in Table 1 show that the here proposed model performs similarly to the original eMoBi-Q. Although, the proposed model achieved a very good prediction performance, indicated by a Pearson correlation ≥ 0.9, further evaluations with other types of distortions are required to draw a more conclusive picture. Nevertheless, the results are an encouraging first step towards an audio quality model for NH and HI listeners.

Loudness predictions as a function of level agree well with measured data as it is shown in panel A of Fig. 2. Panel B in Fig. 2 shows that predictions for the equal loudness contours follow the overall pattern of measured data, however, do not account for the details in the frequency range from 1 kHz to 2 kHz. Although, predictions for spectral loudness summation generally account for increasing loudness as the noise bandwidth increases, some deviations exist in comparison to the data. The effect of spectral loudness summation is underestimated below the reference bandwidth of 3200 Hz and above. Here, a modification of the compression in the peripheral filters or adjustment of parameter in the back end for loudness calculation are potential candidates for a better fit. Overall, the loudness evaluation shows that the newly introduced loudness features capture the relevant effects of loudness perception, which were previously not captured in eMoBi-Q or audio quality models in general. In future developments, such loudness features are intended to be applied as loudness-driven feature maps, where time frames for analyzing distortions are weighted with individual loudness perception (particularly in HI), or are only processed when they exceed a certain loudness.

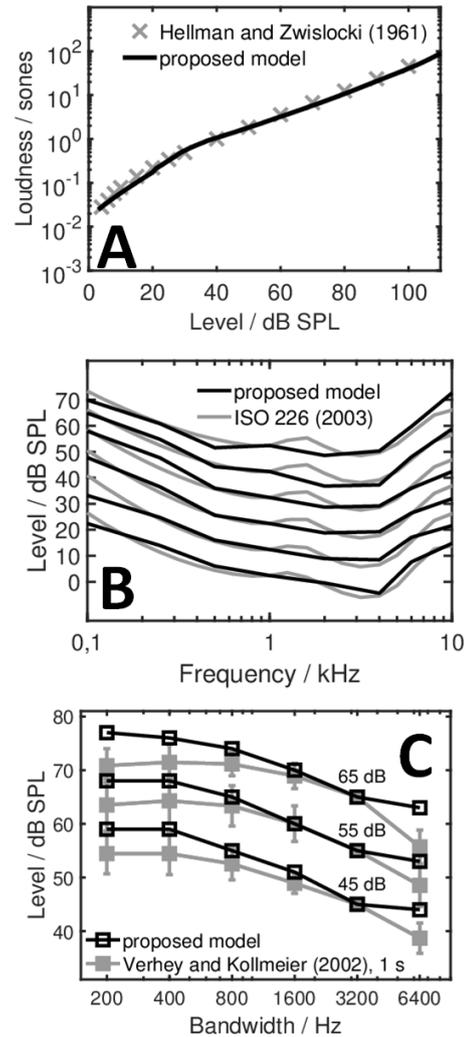

**Figure 2**. Predicted and measured data are shown in black and gray lines (or symbols). Panels A, B, and C refer to loudness as a function of level, equal loudness contours, and spectral loudness summation.

**Table 1.** Pearson correlation coefficients calculated between measured and predicted data are shown for the eMoBi-Q and the proposed model.

| Database | eMoBi-Q | proposed |
|---|---|---|
| Nordholm et al. [11] | 0.99 | 0.97 |
| Fleßner et al. [3] | 0.87 | 0.9 |
| Gößling et al. [12] | 0.98 | 0.99 |
| Schepker et al. [13] | 0.9 | 0.91 |

## 5. CONCLUSIONS

The current study suggests a computationally efficient, combined monaural and binaural audio quality model suited to integrate consequences of sensorineural hearing loss. For this our existing previous approach eMoBi-Q was extended with a non-linear cochlear filterbank, and shown to maintain predictive power for four monaural and binaural audio quality databases. The further extension to predict binaural loudness using the same model front-end offers the prospect of integrating (individually) distorted loudness perception in audio quality predictions for listeners with hearing impairment.

## 6. ACKNOWLEDGEMENTS

This work was supported by the Deutsche Forschungsgemeinschaft (DFG – 352015383 – SFB1330 A2).

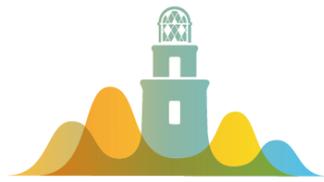